\documentclass[reprint,prx,twocolumn,longbibliography,noeprint,superscriptaddress,floatfix,amssymb]{revtex4-2}

\usepackage{lipsum}
\usepackage[sort&compress]{natbib}
\usepackage{amsmath}
\usepackage{upgreek}
\usepackage{braket}
\usepackage{graphicx}
\usepackage{hhline}

\usepackage[justification=raggedright]{caption}

\usepackage{float}
\usepackage{physics}
\usepackage{graphicx}
\usepackage{xcolor}
\usepackage{amsmath,,amsfonts,amssymb}
\usepackage{textcomp}
\usepackage{bbm}
\usepackage{dsfont}
\usepackage{microtype}
\usepackage{mathtools}
\usepackage{multirow}
\usepackage[utf8]{inputenc}
\usepackage[T1]{fontenc}
\usepackage{hyperref}
\usepackage{caption}
\usepackage{subcaption}

\usepackage{booktabs}

\newcommand\ddfrac[2]{\frac{\displaystyle #1}{\displaystyle #2}}

\newcommand{\fmod}{f_{\textrm{mod}}}
\newcommand{\mus}{\mu\textrm{s}}
\newcommand{\SET}[1]{SET$_{#1}$}

\newcommand{\uu}{$\ket{\uparrow\uparrow}$}
\newcommand{\ud}{$\ket{\uparrow\downarrow}$}
\newcommand{\sdd}{$\ket{\downarrow\downarrow}$}
\newcommand{\du}{$\ket{\downarrow\uparrow}$}
\newcommand{\singFZ}{$\ket{S(4,0)}$}

\newcommand{\uuM}{\uparrow\uparrow}
\newcommand{\udM}{\uparrow\downarrow}

\newcommand{\sddM}{\downarrow\downarrow}
\newcommand{\duM}{\downarrow\uparrow}
\newcommand{\singFZM}{\ket{S(4,0)}}
\newcommand{\Peven}{P$_{\textrm{even}}$}

\newcommand{\Psf}{P_{\textrm{sf}}}

\newcommand{\Pa}{\bar{P}_{r}}
\newcommand{\Pb}{\bar{P}_{l}}

\newcommand{\Pe}{\bar{P}_{e}}
\newcommand{\Pf}{\bar{P}_{f}}
\newcommand{\Pg}{\bar{P}_{g}}

\hypersetup{
    colorlinks,
    citecolor=black,
    filecolor=black,
    linkcolor=black,
    urlcolor=black
}

\begin{document}
\title{Improved Single-Shot Qubit Readout Using Twin RF-SET Charge Correlations}

\author{Santiago Serrano}
\email[]{s.serrano@unsw.edu.au}
\affiliation{
School of Electrical Engineering and Telecommunications,
The University of New South Wales, Sydney, NSW 2052, Australia
}

\author{MengKe Feng}
\affiliation{
School of Electrical Engineering and Telecommunications,
The University of New South Wales, Sydney, NSW 2052, Australia
}

\author{Wee Han Lim}
\affiliation{
School of Electrical Engineering and Telecommunications,
The University of New South Wales, Sydney, NSW 2052, Australia
}
\affiliation{
Diraq, Sydney, NSW 2052, Australia
}

\author{Amanda E. Seedhouse}
\affiliation{
School of Electrical Engineering and Telecommunications,
The University of New South Wales, Sydney, NSW 2052, Australia
}

\author{Tuomo Tanttu}
\affiliation{
School of Electrical Engineering and Telecommunications,
The University of New South Wales, Sydney, NSW 2052, Australia
}
\affiliation{
Diraq, Sydney, NSW 2052, Australia
}

\author{Will Gilbert}
\affiliation{
School of Electrical Engineering and Telecommunications,
The University of New South Wales, Sydney, NSW 2052, Australia
}
\affiliation{
Diraq, Sydney, NSW 2052, Australia
}

\author{Christopher C. Escott}
\affiliation{
School of Electrical Engineering and Telecommunications,
The University of New South Wales, Sydney, NSW 2052, Australia
}
\affiliation{
Diraq, Sydney, NSW 2052, Australia
}

\author{Nikolay V. Abrosimov}
\affiliation{
Leibniz-Institut für Kristallzüchtung, 12489 Berlin, Germany
}

\author{Hans-Joachim Pohl}
\affiliation{
VITCON Projectconsult GmbH, 07745 Jena, Germany
}

\author{Michael L. W. Thewalt}
\affiliation{
Department of Physics, Simon Fraser University, British Columbia V5A 1S6, Canada
}

\author{Fay E. Hudson}
\affiliation{
School of Electrical Engineering and Telecommunications,
The University of New South Wales, Sydney, NSW 2052, Australia
}
\affiliation{
Diraq, Sydney, NSW 2052, Australia
}

\author{Andre Saraiva}
\affiliation{
School of Electrical Engineering and Telecommunications,
The University of New South Wales, Sydney, NSW 2052, Australia
}
\affiliation{
Diraq, Sydney, NSW 2052, Australia
}

\author{Andrew S. Dzurak}
\affiliation{
School of Electrical Engineering and Telecommunications,
The University of New South Wales, Sydney, NSW 2052, Australia
}
\affiliation{
Diraq, Sydney, NSW 2052, Australia
}

\author{Arne Laucht}
\email[]{a.laucht@unsw.edu.au}
\affiliation{
School of Electrical Engineering and Telecommunications,
The University of New South Wales, Sydney, NSW 2052, Australia
}
\affiliation{
Diraq, Sydney, NSW 2052, Australia
}

\begin{abstract}
High fidelity qubit readout is critical in order to obtain the thresholds needed to implement quantum error correction protocols and achieve fault-tolerant quantum computing. Large-scale silicon qubit devices will have densely-packed arrays of quantum dots with multiple charge sensors that are, on average, farther away from the quantum dots, entailing a reduction in readout fidelities. Here, we present a readout technique that enhances the readout fidelity in a linear SiMOS 4-dot array by amplifying correlations between a pair of single-electron transistors, known as a twin SET. By recording and subsequently correlating the twin SET traces as we modulate the dot detuning across a charge transition, we demonstrate a reduction in the charge readout infidelity by over one order of magnitude compared to traditional readout methods. We also study the spin-to-charge conversion errors introduced by the modulation technique, and conclude that faster modulation frequencies avoid relaxation-induced errors without introducing significant spin flip errors, favouring the use of the technique at short integration times. This method not only allows for faster and higher fidelity qubit measurements, but it also enhances the signal corresponding to charge transitions that take place farther away from the sensors, enabling a way to circumvent the reduction in readout fidelities in large arrays of qubits.
\end{abstract}
	
\maketitle

\section{Introduction}\label{sec:Intro}

Electron spins in silicon are promising candidates for large-scale quantum computing \cite{Zwanenburg2013, Saraiva2021} due to their long coherence times \cite{Veldhorst2014, Yoneda2018},  high-temperature operation \cite{Yang2020,  Petit2020, Petit2022,Huang2023_unpublished} and their potential to harness the fabrication capabilities of the semiconductor industry \cite{Gonzalez-Zalba2021}, allowing for densely-packed arrays of qubits. Important advances towards achieving fault-tolerant quantum computing have been achieved by improving the qubit control fidelities \cite{Madzik2022, Xue2022, Noiri2022, Mills2022,Tanttu2023_arxiv}, designing robust multi-qubit addressing techniques \cite{Vahapoglu2021, Vahapoglu2022,Hansen2021,Seedhouse2021_dressed,Hansen2022}, and integrating classical control electronics at the cryogenic level \cite{SchaalCMOS2019, Pauka2021,Xue2021}.

In addition to improved control fidelities, fault-tolerant quantum computing requires high qubit readout fidelities in order to reach the necessary thresholds for quantum error correction \cite{Fowler2012}. Spin-encoded qubits are commonly read using Pauli spin blockade (PSB) to map the spin state of the electron into a dipolar charge movement. Radio-frequency and microwave electrometers, such as radio-frequency single-electron transistors (RF-SETs) \cite{Schoelkopf1998},  gate-based dispersive sensors \cite{Ahmed2018,Pakkiam2018,West2019,Crippa2019,Zheng2019} or single-electron boxes (SEBs) \cite{House2016,Urdampilleta2019,Ciriano-Tejel2020} are particularly suited for fast, high-fidelity spin readout. State of the art reflectometry techniques allow for high readout sensitivities with sub-$\mu$s integration times, and have laid the path for high-bandwidth measurements, significantly above the qubit's decoherence rates \cite{Ibberson2021, Oakes2023, vonHorstig2023_arxiv}. 

Recent progress in quantum dot-based spin qubit devices reveals how an increased number of charge sensors is required as the number of dots per device becomes larger \cite{Hendrickx2021, Borsoi2022,Unseld2023_arxiv,Philips2022}. Consequently, the average separation between the quantum dots and the charge sensors is also increasing, hindering the readout fidelity of large arrays of quantum dots due to the inverse-cube law that governs the field of an electric dipole. A way to circumvent this loss of sensitivity is by indirectly measuring the qubits, either via quantum non-demolition (QND) measurements \cite{Philips2022, Yoneda2020,Takeda2021} or by shuttling the electrons closer to the charge sensors \cite{Li2018}, but these are limited by other sources of infidelity and take longer compared to direct measurements. Previous efforts toward improving the readout capabilities of a pair of SETs (namely, a twin SET) have shown enhanced sensitivities achieved by cross-correlating the individual signals in order to remove local charge noise \cite{Buehler2003, Buehler2004}, but spin qubit readout using correlations from a group of charge sensors is yet to be demonstrated.

\begin{figure*}[ht!]
\centering
\includegraphics[width=\textwidth]{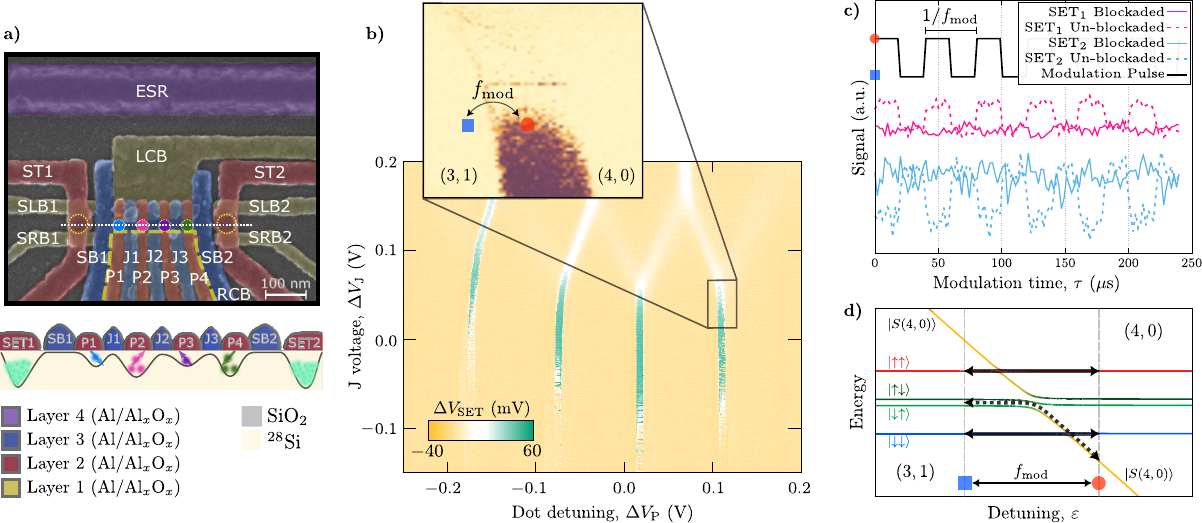}
\caption{\textbf{Device overview and presentation of the modulation technique.} \textbf{a)} False-coloured SEM image of a device nominally identical to the one measured. A cartoon cross-section (below) indicates the layer stack and the accumulation of electrons underneath the gates. \textbf{b)} Charge stability map for a 4-electron configuration under a DQD. The inset shows a Pauli spin blockade (PSB) measurement and the pulsing principle of the modulation technique. A pulse of frequency $f_{\textrm {mod}}$ is applied between (3,1) (blue square) and the PSB window at (4,0) (orange circle). \textbf{c)}~Examples of SET traces (shifted vertically for clarity; 50 averages) as the modulation technique is applied between dots P2 and P3. When the spin system is un-blockaded (blockaded), pulsing into the PSB window back and forth will cause (not cause) periodic tunnelling of the electron between the dots, thus causing the SETs to undergo a corresponding change in signal. \textbf{d)} Energy level diagram of the DQD system. Blockaded spin states [even parity in (3,1)] do not tunnel to (4,0), following the solid paths. Un-blockaded spin states [odd parity in (3,1)] adiabatically tunnel into (4,0), following the dashed path.}
\label{fig:fig1}
\end{figure*}

Here, we expand on the idea of the twin SET and propose a new readout technique that takes advantage of the correlations between a pair of RF-SETs. By modulating the electric dipole signal created by the tunnelling of an electron in a double quantum dot (DQD), we see an improvement in the signal-to-noise ratio (SNR) by ways of filtering out device noise away from the modulation frequency, similar to how a commercial lock-in amplifier works. By recording the modulated time traces instead of a single (integrated) measurement point, we find non-trivial correlations between the independent charge sensors, reaching a reduction in charge readout infidelity of more than one order of magnitude compared to the individual SETs using traditional readout methods. We use a Markov chain model to study the spin decay dynamics associated with the periodic tunnelling of the electron during the readout process and find that relaxation-induced errors can be minimized by operating at high modulation frequencies. This method grants a way to enhance the effective sensitivity of charge sensors, thus providing a path to fast, high-fidelity spin readout in large arrays of qubits. 

\section{Modulation Technique and SET correlations}\label{sec:ModTechnique}

The device measured in this work consists on a linear 4-dot SiMOS device on isotopically enriched silicon (50 ppm $^{29}$Si), where quantum dots are electrostatically defined by aluminium (Al) gates. Fig.~\ref{fig:fig1}a) shows a scanning electron micrograph of a device nominally identical to the one measured, along with a schematic cross section of the device showing the gate stack and the conduction band potential. Critically, the device includes two RF-SETs, one on each end of the dot array. This twin SET configuration enables the detection of charge correlations in order to improve the readout fidelity, as detailed in Section \ref{sec:ChargeReadout}.

The tuning of the device is done by using one of the SETs as an electron reservoir in order to load a specific number of electrons into the quantum dots formed below the P gates. Once the electrons have been loaded, the SET barriers (SLB, SRB, SB) are partially pinched off, decoupling the dots from the reservoir and leaving the total number of electrons constant. This also allows for both SETs to be operated simultaneously as charge sensors. The J gates are used to tune the tunnel coupling between the quantum dots, and the barrier gates (LCB, RCB, SB$_{1}$ and SB$_{2}$) provide additional lateral confinement.

Figure~\ref{fig:fig1}b) shows a typical stability map for a DQD with a total number of 4 electrons loaded. The inset shows a PSB measurement around the $(3,1)\leftrightarrow(4,0)$ transition, with $(N_{L},N_{R})$ the charge occupation in the left and right dot, respectively. Traditionally, spin readout is performed by taking a single measurement at the PSB region (orange circle) and comparing it to a reference (blue square). In this work, we propose extending this method by applying a square modulation pulse (rise time $\sim40\ \text{ns}$) that continuously pulses back and forth between the read (orange circle) and reference (blue square) points at a frequency $f_{\textrm{mod}}$. As shown in Fig.~\ref{fig:fig1}c), the modulation will cause two distinct behaviours on the SET signals depending on the spin state of the electron: when electrons are un-blockaded, modulating back and forth across the charge transition will produce a corresponding square wave on the SET signals due to the activation of the electric dipole associated to the movement of the electron between the two charge configurations. Conversely, when electrons are blockaded, they will remain in the (3,1) charge configuration despite the modulation, and the SET response will not change [see Fig.~\ref{fig:fig1}d)]. Demodulating the full SET traces at $f_{\textrm{mod}}$ will result in separate scalar values for the blockaded and un-blockaded states.  The pulsing rates and tunnel couplings employed in this work are such that the PSB readout does not differentiate between singlet and triplet T$_{0}$ states, resulting in a parity readout where the even-parity ($\ket{\downarrow\downarrow},\ \ket{\uparrow\uparrow}$) states are blockaded and the odd-parity ($\ket{\downarrow\uparrow},\ \ket{\uparrow\downarrow}$) states are un-blockaded \cite{Seedhouse2021}. 

It is worth noting that the SET responses are composed of two contributions: the periodic signal from the modulated electron tunnelling and noise.  The un-blockaded tunnelling signal measured by SET$_{1}$ is anti-correlated to that measured by SET$_{2}$ (given that both SETs are tuned on the same slope of the Coulomb peak) as an electron moving closer to SET$_{1}$ moves farther from SET$_{2}$. On the other hand, noise, which we assume to be local due to the spatial distribution of two-level fluctuators (TLFs) in the device, is uncorrelated. In the following section, we show how these correlations can be exploited to improve the readout fidelity.

\section{Improved Charge Readout}\label{sec:ChargeReadout}

\begin{figure*}[t!]
\centering
\includegraphics[width=\textwidth]{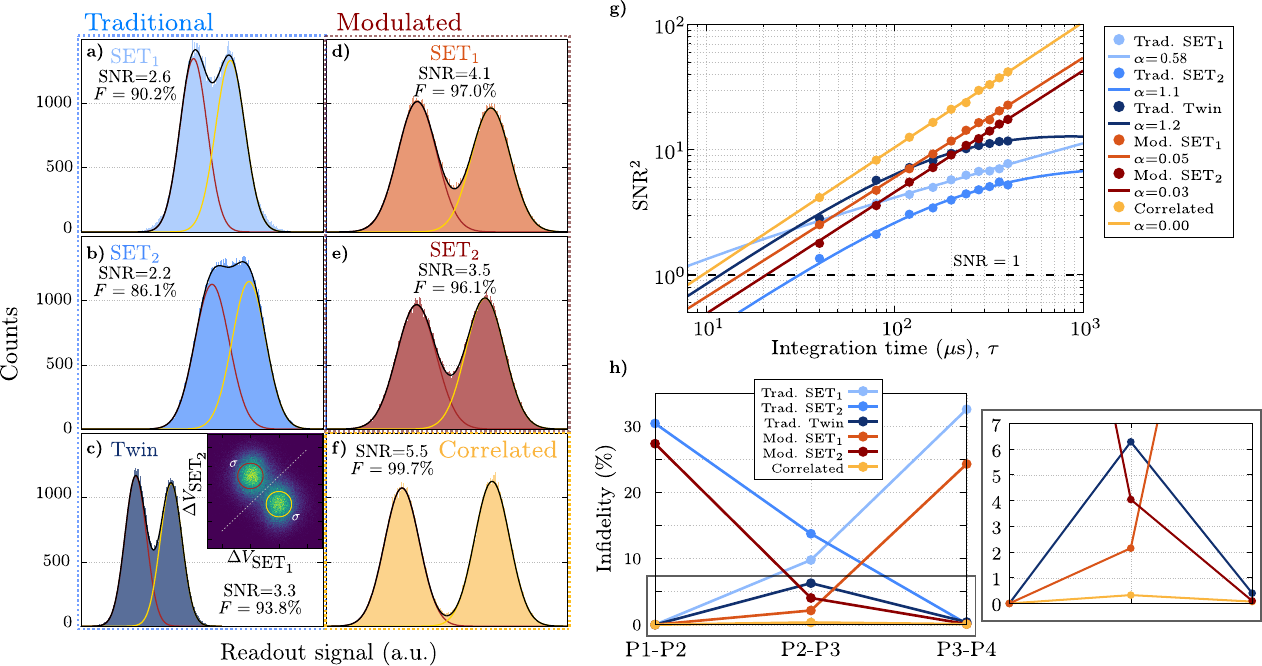}
\caption{\textbf{Improved charge readout fidelity.} \textbf{a)-f)} Single-shot charge readout histograms for the P2-P3 transition obtained using different techniques. The total integration time for each technique is $\tau=$ 280 $\mu$s per shot. \textbf{a),~b)} Traditional baseband readout method using single SET$_{1}$ and single SET$_{2}$, respectively. \textbf{c)} Traditional baseband readout method using a twin SET. The inset shows the 2D histogram for the twin SET. \textbf{d),~e)} Modulated readout method using single SET$_{1}$ and single SET$_{2}$, respectively.  $f_{\textrm{mod}}=$ 25 kHz. \textbf{f)} Correlated readout method using the twin SET modulated signal. $f_{\textrm{mod}}=$ 25 kHz and the correlation function is given in Eq. \ref{eq:corr}. \textbf{g)}~Square of the SNR as a function of integration time for the different techniques.  Solid lines plot the fits to the function $\textrm{SNR}^{2} = S_{0}\tau/ \left(1+A_{f}\tau^{\alpha}\right)$. \textbf{h)} Impact of the location of the charge transition on the charge readout infidelity for the different techniques. $\tau=280\ \mus$, $\fmod=25\ \text{kHz}$. Solid lines are used as guides to follow the trend of the points. The inset is a zoom-in to the region indicated by the box.}
\label{fig:fig2}
\end{figure*}

Having introduced the idea of the modulation technique, we proceed to study its performance in comparison with the traditional, single-point, single SET technique. To do so, we initialize a single electron in the P2-P3 DQD system and perform the modulation technique at two different points: close to the $(0,1)\leftrightarrow(1,0)$ transition, such that the modulation causes the electron to tunnel back and forth between the dots; and deep inside $(0,1)$, so that the charge configuration does not change with the pulsing. This way, we are able to assess the quality of the pure charge readout process without any spin-to-charge conversion infidelity effects. A single-shot total integration time $\tau=$ 280 $\mu$s was used in all cases. For the traditional measurements, this means that the integration time at each of the read and reference points was 140 $\mu$s. For the modulated measurements, the total modulation time was 280 $\mu$s; these measurements require no reference.

Figures~\ref{fig:fig2}a--f) show the charge readout histograms obtained using different readout techniques. The SNR is limited by the distance from the P2-P3 DQD system to the SETs, as well as the screening of the aluminium gates in-between. Figures~\ref{fig:fig2}a,b) show the results for the traditional measurement using SET$_{1}$ and SET$_{2}$, respectively.  The advantage of the twin SET is evident once we combine the independent SET signals in a way that improves the fidelity of the readout, as seen in Fig.~\ref{fig:fig2}c), corresponding to the linear combination obtained by projecting the data along the axis that maximizes the separation between the readout peaks (see inset and Appendix~\ref{sec:twinSET_corrFunct}).

The readout results using the modulation technique on SETs 1 and 2 are shown in Figs.~\ref{fig:fig2}d,e), respectively. The data plotted in the histograms corresponds to the scalar result obtained after demodulating the SET traces at the modulation frequency, $f_{\textrm{mod}}=25$ kHz, using sinusoidal tones. A considerable improvement in the SNR and, consequently, the charge readout fidelity, is observed in comparison to their traditional baseband counterparts. This improvement is a direct consequence of the modulation/demodulation, which effectively acts as a  single-shot lock-in measurement with a bandpass filter around $f_{\textrm{mod}}$, thus protecting the signal from noise at the device level.

In addition to providing noise immunity, the modulation of the charge sensing signal opens a path to exploiting more sophisticated correlations between the twin SET traces. The correlation function used in this work is given by:
\begin{eqnarray}
&\mathcal{C}=\mathcal{D}\Big\lbrace\left[\text{SET}_{1}(t)-a\right]\cdot\left[\text{SET}_{2}(t)+a\right] + a^{2}\Big\rbrace \nonumber\\
&\mathcal{C}=\mathcal{D}\Big\lbrace \underbrace{\text{SET}_{1}(t)\cdot \text{SET}_{2}(t)}_{\Pi}+a\underbrace{\left(\text{SET}_{1}(t)-\text{SET}_{2}(t)\right)}_{\Delta}\Big\rbrace \nonumber\\
&\mathcal{C}=\mathcal{D}\Big\lbrace \Pi + a\Delta\Big\rbrace, \label{eq:corr}
\end{eqnarray}
where $\text{SET}_{1}(t)$ and $\text{SET}_{2}(t)$ are the modulated traces for SET$_{1}$ and SET$_{2}$, respectively, $\mathcal{D}$ is the demodulation function $\mathcal{D}\lbrace f(t)\rbrace=\int_{0}^{\tau}f(t)\sin(2\pi f_{\textrm{mod}}t)dt$, and $a$ is a free parameter that tunes the relative weight between the product, $\Pi$, and the difference, $\Delta$, of the traces.  This correlation function significantly enhances the SNR because the difference, $\Delta$, amplifies the anti-correlations between the traces (signal), while the product, $\Pi$, cancels out uncorrelated elements (noise). In addition, the demodulation provides an extra layer of immunity to noise by filtering out frequency components away from $f_{\textrm{mod}}$.  The single-shot histogram corresponding to the correlation between the SET traces is shown in Fig.~\ref{fig:fig2}f), and it exhibits a reduction of more than one order of magnitude in the charge readout infidelity when compared to the traditional, single SET technique. It is worth noting that the correlation between the traces is performed shot-by-shot without taking any averages.

In order to further characterize the improvement of the readout, we repeat the same measurement for different values of total integration time, $\tau$, the results of which are plotted in Fig.~\ref{fig:fig2}g). The fitting of the data to the function $\textrm{SNR}^{2} = S_{0}\tau/ \left(1+A_{f}\tau^{\alpha}\right)$ shows that the results that involve modulation have a dependency close to SNR~$\propto\sqrt{\tau}$ ($\alpha\approx0$), expected of measurements limited by white noise. Conversely, the larger values of $\alpha$ obtained for the traditional measurements indicate a slower increase of SNR with integration time, consistent with signals affected by $1/f^{\alpha}$ (also known as ``pink'' or ``flicker'' for $0<\alpha<2$) noise \cite{McDowell2007, McDowell2008} (see Appendix~\ref{sec:ImmunityToPinkNoise}). In the regime where pink noise is dominating, longer integration times will begin to capture noise signals that are higher in amplitude, thus losing the benefit of averaging over an increased number of samples. On the other hand, the modulation acts as a bandpass filter around $f_{\textrm{mod}}$, rendering the measurement insensitive to pink noise. This result evidences the immunity to electrical noise provided by the modulation, and the further improvement in readout sensitivity enabled by the correlation of the two signals.

An additional advantage of using a twin SET is that the reduction in signal that comes from moving the sensors away from the qubits can be compensated. In Fig.~\ref{fig:fig2}h), we plot the charge readout infidelity obtained with the different methods as a function of the double-dot pair, which is equivalent to changing the distance between the center of the electric dipole and the sensors. For comparison purposes, the total integration time in all cases is kept constant at 280 $\mu$s and the modulation frequency is $\fmod=25\ \text{kHz}$. As expected, transitions between P1-P2 (P3-P4) are more weakly detected by SET$_{2}$ (SET$_{1}$) and the readout infidelity of the single SET measurement increases as the dipole is moved away from the sensor. Notably, this monotonic trend is avoided by the twin SET, with the correlation function providing below 0.3\% readout infidelity ($>99.7\%$ readout fidelity) throughout the array.

\section{Modulated Spin Readout}\label{sec:SpinModulation}

\begin{figure*}[ht!]
\centering
\includegraphics[width=\textwidth]{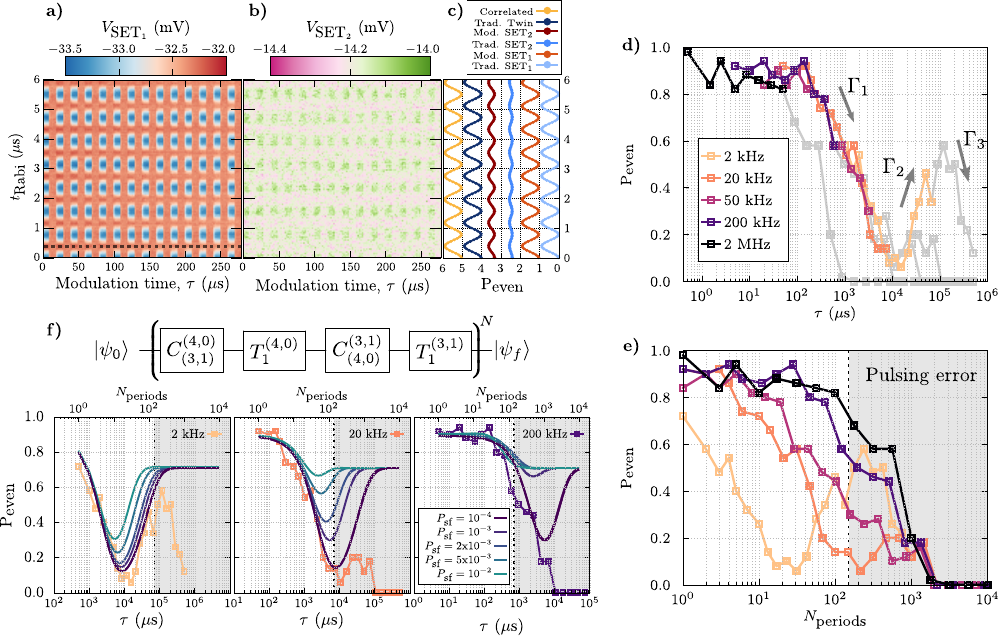}
\caption{\textbf{Modulated spin qubit readout.} \textbf{a)} SET$_{1}$ and \textbf{b)} SET$_{2}$ time traces vs microwave pulse duration as we apply the modulation technique with $\fmod=50\ \textrm{kHz}$.  Spin rotations change the parity of the 2-qubit (P1-P2) system, transitioning between blockaded and un-blockaded states. \textbf{c)} Single-shot results for the different techniques showing Rabi oscillations (data offset for clarity). \textbf{d)} Probability of even parity measured by \SET1 at the end of the modulation technique along the dashed line in a) for different modulation frequencies. The traces are colour-coded up to 150 modulation periods and continue in grey colour afterwards to indicate the point where the systematic pulsing error becomes significant. \textbf{e)} Same data as in d) replotted as a function of the number of modulation periods. \textbf{f)}~A Markov chain consisting of crossing ($C$) and relaxation ($T_{1}$) processes is used to model the evolution of the spin system. Each of the bottom panels shows the experimental data at a given modulation frequency (squares) and the theoretical fitting for different values of the probability of crossing-induced spin flip, $P_{\textrm{sf}}$,  (solid lines) assuming an \uu\ initialization. The shaded area after 150 modulation periods in e) and f) highlights the data points affected by the systematic pulsing error. }
\label{fig:fig3}
\end{figure*}

We now study the impact that modulating the electron back and forth across a charge transition during readout has on its spin degree of freedom. We load four electrons into the P1-P2 DQD subsystem in the (3,1) charge configuration [see Fig.~\ref{fig:fig1}b)]. Here, the lower two electrons under P1 form a spin-zero closed shell,  leaving the remaining two electrons for us to use as spin qubits \cite{Yang2020}. We measure the even (blockaded) state probability of the 2-qubit system as we sweep the duration, $t_{\textrm{Rabi}}$, of a microwave field with a frequency resonant with the spin's Larmor frequency, i.e., $f_{\textrm{MW}}=g\mu_{\textrm{B}}B_{0}/h$. Here, $g$ is the electron $g$-factor, $\mu_{\textrm{B}}$ is the Bohr magneton, $B_{0}\approx700$ mT is the external DC magnetic field that defines the spin quantization axis, and $h$ is Planck's constant.  We initialize the 2-qubit system by ramping from (4,0) into (3,1) at a rate that results in an un-blockaded (odd) state and choose $f_{\textrm{MW}}$ to be on resonance with the lower-frequency spin \cite{Huang2019}.  The microwave field causes the resonant spin to coherently oscillate between $\ket{\downarrow}$ and $\ket{\uparrow}$, thus periodically changing the parity of the 2-qubit system between even and odd. We perform modulated measurements at the PSB point using both SETs simultaneously, and interleave them with traditional measurements in order to benchmark the spin readout visibility.

Figures~\ref{fig:fig3}a, b) show the results of the modulated measurements on SET$_{1}$ and SET$_{2}$, respectively, before performing the demodulation. Oscillations along the $x$-axis are due to the modulation pulse, while oscillations along the $y$-axis correspond to coherent qubit control. When the spins are in an odd (un-blockaded) state, the electron is free to tunnel back and forth and the SETs exhibit corresponding signals. Conversely, after applying a $\pi$-rotation to one of the qubits to produce an even (blockaded) spin state, tunnelling is forbidden and the SETs' signals remain constant. In Fig.~\ref{fig:fig3}c), we compare the Rabi oscillations obtained by the different techniques (i.e., after demodulation), demonstrating that our proposed technique can be used for single-shot qubit readout.

In the case of SET$_{1}$, the Rabi visibility is limited by initialization and spin-to-charge conversion to about $86\%$, even in the case of the traditional readout, and the modulation/correlation does not exhibit a noticeable improvement. However, the increased distance to the P1-P2 electric dipole entails that the visibility measured by SET$_{2}$ is limited by the charge readout fidelity, and an important improvement from $14\%$ to $30\%$ is obtained by the modulation. This factor of 2 improvement in qubit visibility is a direct consequence of the SNR improvement proper of the technique. Notably, by correlating the SETs the visibility of SET$_{1}$ is maintained as the correlation function filters out the predominant uncorrelated noise from SET$_{2}$.

Next, we explore how the modulation frequency, $f_{\textrm{mod}}$, and total modulation time, $\tau$, affect the decay of the spin states. To avoid sampling limitations (see Appendix~\ref{sec:limitations}) we no longer record the SET traces as we modulate across the transition. Instead, we apply the modulation technique for different values of $f_{\textrm{mod}}$ and $\tau$, and perform a single traditional measurement with \SET1 (260 $\mus$ integration time) at the end. As with the Rabi measurements, we initialize the system in an un-blockaded state. We then apply a control $\pi$-rotation on the lowest energy qubit, resulting in an even-parity (blockaded) state. We also perform a control experiment in which we set the amplitude of the modulation pulse to zero so that it causes no tunnelling across the charge transition, constituting a $T_{1}$ measurement inside (3,1). The results are plotted in Figs.~\ref{fig:fig3}d), e).

We point out that it is unclear to us whether the ramp into (3,1) to initialize the un-blockaded state results in \ud, \du, or a mixed state between the two, although the behaviour of the spin decay suggests a predominant \ud\ initialization (see Appendix~\ref{sec:spinInit} for further comments on the spin initialization). Irrespective of this unknown, the structure of the spin-to-charge conversion process along with the observed coherent driving are sufficient to study the effect of the modulation technique on the qubits and draw the conclusions presented here.

The evolution of the even probability in Fig.~\ref{fig:fig3}d) consists of three distinct transitions. The first decay in \Peven\ ($\Gamma_{1}$) corresponds to the $T_{1}$ relaxation at one of the dots. This relaxation occurs both inside (3,1) and (4,0), where the resulting states are, respectively,  an odd state (\du, \ud) and \singFZ. The subsequent rise in \Peven\ ($\Gamma_{2}$) is the $T_{1}$ relaxation at the other dot, leading to \sdd. This relaxation can only occur inside (3,1) since odd states are directly mapped into \singFZ\ when crossing into (4,0). The final transition ($\Gamma_{3}$) corresponds to a systematic pulsing error in the experiment, by which the detuning level was being shifted slightly with every modulation period, eventually causing the pulsing to take place entirely inside (4,0) and moving the read point out of the PSB window. This resulted in a suppression of \Peven\ after a fixed number of periods independent of the modulation frequency, as evidenced in Fig.~\ref{fig:fig3}e). We have taken a cautious approach and have marked the data consisting of more than 150 modulation periods (shown in grey) as likely affected by this systematic error. 

We confirm that $\Gamma_{1}$ and $\Gamma_{2}$ correspond to the spin relaxation processes described above since they are frequency-independent phenomena that occur at a specific total time irrespective of the number of periods [see Figs.~\ref{fig:fig3}d),~e)] and match the relaxation rates observed in the control experiment where there was no pulsing across the charge transition (see Appendix~\ref{sec:markovChain} and Fig.~\ref{fig:markovControl}). The fact that the repeated crossing induces no significant spin errors favours the use of the readout technique at high modulation frequencies, where a larger number of periods can be reached within shorter integration times.

To describe the spin dynamics undergone during the modulation, we use a Markovian process consisting of four independent contributions: the crossing from $(3,1)$ to $(4,0)$, $C_{3,1}^{4,0}$; the wait time at $(4,0)$, $T_{1}^{4,0}$; the crossing from $(4,0)$ back to $(3,1)$, $C_{4,0}^{3,1}$; and the wait time at $(3,1)$, $T_{1}^{3,1}$. The crossing processes induce spin flip errors with probability $\Psf$ and the wait times allow the system to inelastically relax to a lower-energy state (see Appendix~\ref{sec:decayAccum} and \ref{sec:markovChain} for more information). After $N$ modulation periods, the operator describing the evolution of the system is:
\begin{eqnarray}
O = \left[T_{1}^{3,1} \cdot C_{4,0}^{3,1} \cdot T_{1}^{4,0} \cdot C_{3,1}^{4,0}\right]^{N},
\end{eqnarray}
where $\cdot$ denotes matrix multiplication. As evidenced in Fig.~\ref{fig:fig3}f), our model correctly describes the frequency-independent relaxation processes (assuming an initial \uu\ state after the $\pi$-rotation, see Appendix~\ref{sec:spinInit}), with the best fit obtained when we assign the probability of crossing-induced spin flip errors to $\Psf=10^{-4}$, supporting the conclusion that spin errors are dominated by relaxation rather than the crossing itself.

\section{Discussion}

\subsection{Modulation-Induced Errors and High-Frequency Operation}

The method we present here allows for significant improvements in charge readout fidelities as evidenced in Fig.~\ref{fig:fig2}. However, it is important to consider that the periodic pulsing of the electron between the two charge configurations gives way to relaxation-induced errors. By spending time inside (3,1) and (4,0), states tend to relax to \sdd\ and \singFZ, respectively, thus reducing the spin readout fidelity.  It becomes necessary to minimize the time spent on each side whilst still pulsing for a sufficiently large number of periods so as to not penalize the demodulation (see Appendix~\ref{sec:modTechCharact}). The results from Figs.~\ref{fig:fig3}d),~e) provide evidence that this condition is met by operating at high frequencies, which allows multiple crossings to take place during short total times, limiting the $T_{1}$ relaxation as more periods are required to accumulate enough probability to decay (see Appendix~\ref{sec:decayAccum}). On the other hand, too high modulation frequencies can lead to a regime where errors are dominated by crossing-induced spin flips. The non-zero probability $\Psf$ can quickly accumulate as the frequency increases, resulting in limited spin readout fidelity even over short integration times, although spin fidelities have been demonstrated to be maintained after thousands of inter-dot crossings \cite{Yoneda2021}.

\subsection{Generalization to Other Platforms}\label{sec:generalization}

\begin{figure}[ht!]
\centering
\includegraphics[width=0.45\textwidth]{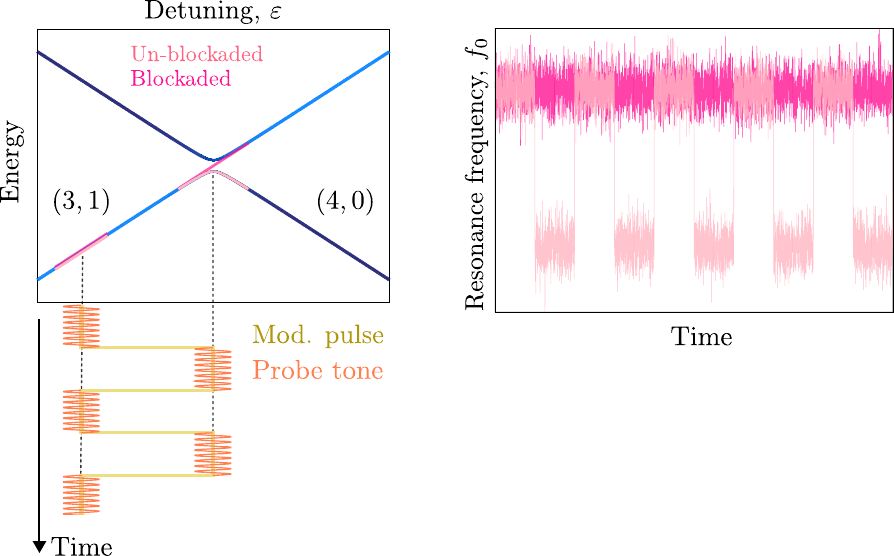}
\caption{\textbf{Generalization to gate-based sensors.} Schematic of a way to apply the modulation technique to gate-based dispersive readout. A detuning pulse modulates the frequency of the resonator by periodically probing the quantum capacitance at two different points. }
\label{fig:general2DRO}
\end{figure}

The methods presented in this work are not exclusive to an RF-SET-based sensing scheme, and other high-bandwidth techniques such as gate-based dispersive readout and single-electron boxes can benefit from the modulation and correlation of signals from a group of sensors. One way in which we envision our technique being applied to dispersive readout is by using the detuning pulse to probe the quantum capacitance of the system at a reference point [i.e., deep inside (3,1)] and compare it to that at the charge transition $(\varepsilon=0)$. Depending on the spin state of the electrons, the capacitance at the charge transition will change and the frequency of the resonator will be modulated accordingly. Fig.~\ref{fig:general2DRO} depicts a diagram of how this implementation would work. It is not unrealistic to contemplate a distributed array of gate-based sensors, each capacitively coupled to a number of gates, in which cross-correlations between the independent modulated traces are exploited to improve the joint sensitivity of the system.

\section{Conclusions and Outlook}

We have presented a technique that enhances the sensitivity of charge sensors in two ways: first, by modulating the sensor signal at a frequency $\fmod$ in order to grant immunity to charge noise containing frequency components that are not in proximity to $\fmod$; and second, by combining two independent modulated signals in order to enhance correlations and suppress uncorrelated noise.  The method shows improvement in charge readout infidelities of over one order of magnitude, and can be used to readout spin qubits with high visibility. In the cases where the qubit visibility is limited only by charge readout, this technique can enable significant improvements in qubit readout fidelity without any cost in integration time. Similarly, it provides a way to readout distant dots, effectively reducing the amount of sensors needed in a large dot array.

We anticipate that this technique can be extended to a wide range of high-bandwidth sensors, including SEBs and gate-based dispersive sensors. In essence, any set of signals that can be modulated in time can be correlated to enhance their collective sensitivity in a meaningful manner.

This technique benefits from high modulation frequencies (limited only by the interdot tunnel coupling, typically of several GHz, and the measurement bandwidth) and fast sampling rates in order to avoid aliasing effects. For the integration times studied, we found that spin flips induced by the periodic tunnelling across the charge transition are not the dominant source of errors, and time-dependent relaxation processes govern the spin decay. It is therefore desirable to use hundreds of periods at large modulation frequencies and short integration times, thus mitigating the relaxation-induced spin errors and, simultaneously, improving the quality of the demodulation.

Finally, more sophisticated correlation functions can be studied in order to further improve the joint sensitivity of the sensor array.

\section*{Acknowledgements}

We acknowledge support from the Australian Research Council (FL190100167 and CE170100012), the US Army Research Office (W911NF-23-10092) and the NSW Node of the Australian National Fabrication Facility. The views and conclusions contained in this document are those of the authors and should not be interpreted as representing the official policies, either expressed or implied, of the Army Research Office or the US Government. S.S., M.K.F. and A.E.S acknowledge support from Sydney Quantum Academy. 

\section*{Author Contributions}
S.S. performed the experiments and analysed the data. M.K.F. developed the theoretical model to analyse the spin dynamics. W.H.L. and F.E.H. fabricated the device. C.C.E. contributed to the design of the device. M.L.W.T., N.V.A. and H.-J.P. prepared and characterized the isotopically purified silicon substrate. A.E.S. contributed to the formulation of the correlation technique.  S.S and W.G. wrote the measurement code. T.T., A.S. and A.L. contributed to discussions on experimental results and data analysis.  S.S. wrote the manuscript with input from all co-authors. A.S.D. and A.L. supervised the project.

\appendix

\section{Measurement Setup}
The device is measured in a Bluefors LD400 dilution refrigerator operated at a base temperature of 14 mK.

D.C. biasing of the device is done using a QDevil QDAC. Gate pulsing waveforms are generated with a Quantum Machines OPX+, and are combined with the D.C. voltages using linear bias-tees at room temperature. The microwave signals for spin control are generated using a Keysight PSG8267D vector signal generator, with IQ (single-sideband) and pulse modulation waveforms generated by the OPX+.

All the signals are filtered at the mixing chamber using custom-made low-pass filters at 30 Hz for the D.C. lines and 400 MHz for the pulsed lines.

Both SETs used in this work are configured for reflectometry-based measurements (RF-SET). A surface mount inductor is connected to the drain lead and a 100~pF grounding capacitor to the source. The values of inductance used are $L_{1}=750$ nH and $L_{2}=680$ nH, which result in resonance frequencies of $f_{1}=180$ MHz and $f_{2}=210$ MHz for RF-SET$_{1}$ and RF-SET$_{2}$, respectively.  

Two independent amplifying chains, one for each RF-SET,  are used.  A cryogenic amplifier, LNF-LNC0.2\_3A, is located at the 4 K (still) stage of the dilution refrigerator for RF-SET$_{1}$ (RF-SET$_{2}$),  followed by a series of room temperature amplifiers, MiniCircuits ZX60-P103LN+ and MiniCircuits ZFL-1000. Directional couplers, MiniCircuits ZX30-12-4, mounted on the mixing chamber plate of the dilution refrigerator are used to separate the signal coming into the device from the reflected one that goes into the amplifying chains.

The generation of the RF tones, the sampling of the reflected RF signals and their subsequent (primary) demodulation at the RF frequency is performed using the Quantum Machines OPX+.

\section{Real-Time Correlations}
All the correlations presented in this work were calculated numerically by post-processing the recorded SET traces. Likewise, the (secondary) demodulation of the SET traces at $f_{\textrm{mod}}$ was performed numerically during post-processing. Nevertheless, all the calculations used here require only arithmetic functions that can be efficiently implemented using real-time hardware operations such that the method can be fully implemented on-the-fly in a fault-tolerant architecture.

\section{Limitations}\label{sec:limitations}

We find that the biggest limitation on our implementation of the technique comes from hardware constraints which prevented us from using sampling frequencies above 500 kSa/s. The on-board FPGA of the QM OPX+, which we use to apply the modulation pulse and digitize the SET traces, has a limited program memory, and resource allocation criteria meant that the compilation of the program would fail when trying to sample above 500 kSa/s. This limitation can be easily overcome by more efficiently using the FPGA resources, or by using commercially available signal generators and digitizers whose characteristics comfortably allow for operation above tens of MHz.

According to Nyquist's sampling theorem, the sampling frequency determines an upper boundary for $f_{\textrm{mod}}$. Since the modulation frequency sets the number of periods of the signal for a given integration time, accessing higher modulation frequencies (above MHz) would enable higher sensitivity at lower integration times because the readout would not be limited by the demodulation of few-period signals (see Fig.~\ref{fig:snr_vs_fmod_P2P3} and Fig.~\ref{fig:compareHistsPeriods}). We point out that reflectometry-based readout is particularly suited for this purpose due to the high bandwidths that it enables.

During the processing of the data, a systematic error was found whereby the finite resolution of the arbitrary waveform generator (AWG) used to generate the modulation pulses caused an accumulated voltage drift of $2^{-16}\approx15.3\ \mu\textrm{V}$ (one least significant bit) per modulation period. After about 150 periods, this accumulated error was significant enough to change the characteristics of the spin relaxation processes and the accuracy of the readout for the spin decay experiments presented in Sec.~\ref{sec:SpinModulation}.

A more fundamental limitation of the technique relates to the tunnel rate of the DQD interdot charge transition. The anti-correlated signal between both SETs is caused by the dipolar transition that originates when the electron tunnels back and forth between the two dots. Naturally, this requires that the electron tunnels within the modulation period, so tunnel rates below $\fmod$ will cause latching that will be penalized by the demodulation function. As shown in Fig.~\ref{fig:ModCharacterization_J_modAmp}, the readout signal drops when the tunnel rate falls below a certain threshold, given by the modulation frequency.

\section{PSB Measurement}
The pulsing sequence used to obtain the PSB maps like the one shown in the inset of Fig.~\ref{fig:fig1}b) consists of initializing two different spin states and comparing the readout signals \cite{Johnson2005}. We first initialize a $\ket{S(4,0)}$ by pulsing into the (4,0) charge configuration and waiting for about $100\ \mu$s to allow for spin relaxation to take place. We then sweep the readout point around the charge transition and perform a first measurement (read$_{\textrm{Sing}}$). Subsequently, we initialize a mixed state between $\ket{\uparrow\downarrow}$, $\ket{\downarrow\uparrow}$ and $\ket{\downarrow\downarrow}$ by pulsing from (4,0) to (3,1) at a rate of about $100\ \mu\text{V}/\mu\text{s}$, followed by a second readout at the same readout point as before (read$_{\textrm{Mix}}$). The PSB map shows the difference read$_{\textrm{Sing}}$ - read$_{\textrm{Mix}}$.

\section{Spin Initialization}\label{sec:spinInit}

Based solely on the behaviour of the spin relaxation from Figs.~\ref{fig:fig3}d), \ref{fig:fig3}e) and Fig.~\ref{fig:markovControl}, it seems clear that the state of the 2-qubit system at the beginning of the modulation technique (i.e., following the $\pi$-rotation on the lowest-energy spin) is \uu. This follows from the fact that the double relaxation ($\Gamma_{1}$ decaying down to \Peven$<0.1$, followed by $\Gamma_{2}$) can only be explained if the state at the beginning of the experiment is predominantly \uu, since any \sdd\ population will remain blockaded, setting a lower limit for \Peven. It is possible for \sdd\ to relax into \singFZ\ at (4,0) so that \Peven\ goes below this lower limit, but this process would need to be faster than $\Gamma_{2}$ and the rise in \Peven\ after $10^{4}\ \mus$ would not take place. In other words, the fact that \Peven\ goes to almost zero and then increases again requires very little \sdd\ population at the beginning of the modulation.

Since the \uu\ state is obtained by applying a $\pi$-rotation on the lowest energy spin, the initial spin state after the ramp into (3,1) must be \ud. However, in a standard 5-level spin system \cite{Fogarty2018} a single ramp across the charge transition will seldom result in a predominant \ud\ initialization because state mixing occurs across the anti-crossing. Alternatively, a high \uu\ population can be obtained if the starting ramp adiabatically crosses into a pure \du\ state, after which the $\pi$-rotation is applied on the \textit{highest} energy qubit, indicating that the single-sideband modulation used for the control pulse is inverted; no other evidence of this inversion was found.

Therefore, and as stated in the main text, we remain unsure about the initial spin state of the system. Nevertheless, we believe that this does not impede or change our assessment of the spin decay.

\section{Immunity to \texorpdfstring{$1/f$}{1/f} Noise} \label{sec:ImmunityToPinkNoise}
The results from Fig.~\ref{fig:fig2}g) show that the modulation technique provides immunity to flicker noise by acting as a bandpass filter around the modulation frequency $f_{\textrm{mod}}$. By fitting the data to the function $\textrm{SNR}^{2} = S_{0}\tau/ \left(1+A_{f}\tau^{\alpha}\right)$, we can identify the effect that different sources of noise have on the measurement. This function describes signals which have different dominant sources of noise depending on the time scale, and provides a good way to account for the transition between white noise-dominated signals at low $\tau$ and $1/f$ noise-dominated at large $\tau$ \cite{McDowell2007, McDowell2008}. $S_{0}$ is the overall amplitude level of the signal, $A_{f}$ sets the time scale for the change in regime between white and $1/f$ noise, and $0<\alpha<2$ corresponds to the exponent of the flicker noise, $1/f^{\alpha}$, and it is associated to the distribution of two-level charge fluctuators that couple to the measured signal. The values of the fitting parameters are registered in Table \ref{table:SNRvsTimeFitting}.

\begin{table}[h!]
\centering
\caption[Fitting parameters.]{Fitting parameters for the measurements from Fig.~\ref{fig:fig2} to the function $\textrm{SNR}^{2} = S_{0}\tau/ \left(1+A_{f}\tau^{\alpha}\right)$. The value of $\alpha$ corresponds to the distribution of two-level fluctuators with $1/f^{\alpha}$ dependency. The error bars indicate the $95\%$ confidence intervals.}
\begin{tabular}{cccc}\toprule
Readout technique & $\alpha$ & $S_{0}$ & $A_{f}$\\ \midrule
Trad. SET$_{1}$ & 0.58 $\pm\ 0.04$ & 1.0 $\pm\ 0.1$ & 1.6 $\pm\ 0.3$\\  
Trad. SET$_{2}$ & 1.1 $\pm\ 0.6$ & 0.03 $\pm\ 0.02$ & $0.002 \pm 0.009$\\  
Trad. Twin & 1.2 $\pm\ 0.2$ & 0.09 $\pm\ 0.01$ & $0.001 \pm 0.002$ \\  
Mod. SET$_{1}$ & 0.05 $\pm\ 0.06$ & 1.0 $\pm\ 0.1$ & 12 $\pm\ 5$\\  
Mod. SET$_{2}$ & 0.03 $\pm\ 0.03$ & 1.0 $\pm\ 0.1$ & 18 $\pm\ 3$\\  
Correlated & 0.00 $\pm\ 0.05$ & 1.0 $\pm\ 0.1$ & 9 $\pm\ 3$ \\ \bottomrule
\end{tabular}
\label{table:SNRvsTimeFitting}
\end{table}

We note that even though we use RF-SETs, which by themselves have a layer of protection against $1/f$ noise due to the modulation of the drain Fermi level, charge noise can still affect the readout performance. The RF-SET is only protected against $1/f$ noise on the measurement setup, including ground loops and the amplifying chain up to the digitizer. However, spurious charge fluctuations at the device level will modulate the amplitude of the RF-SET signal and, upon demodulation, will be indistinguishable from true charge sensing signals. The secondary modulation that we propose provides protection against charge noise at the device level, since any processes that affect the sensor's conductance and may pass as signal -- such as a two-level fluctuator strongly coupled to the SET -- will be filtered out unless they occur at the known frequency $\fmod$.

\section{Twin SET and Correlation Functions}\label{sec:twinSET_corrFunct}
\begin{figure}[ht!]
\centering
\begin{subfigure}[h]{0.25\textwidth}
\includegraphics[width=\textwidth]{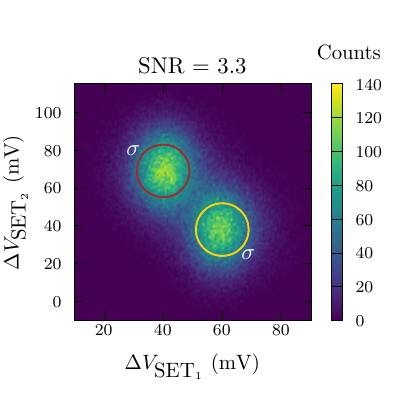}
\caption{}
\label{fig:2D_hist_twin}
\end{subfigure}%
\begin{subfigure}[h]{0.25\textwidth}
\includegraphics[width=\textwidth]{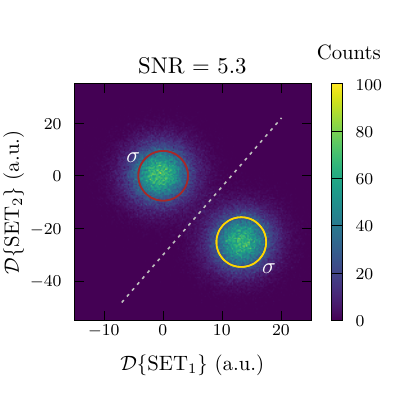}
\caption{}
\label{fig:2D_hist_diff_mod}
\end{subfigure}%
\caption{Twin SET 2D histogram for the traditional (a) and modulated (b) readout. The projection along the dashed axis maximizes the SNR.}
\label{fig:2D_hist}
\end{figure}

The independent SET measurements can be treated as quadrature signals of a single measurement, which results in a 2D histogram as seen in Fig.~\ref{fig:2D_hist_twin}. The single-shot measurement for the Twin SET is obtained by finding the axis on this histogram that maximizes the separation between the readout peaks and projecting the data along that axis. For the particular case of the measurement described in Sec.~\ref{sec:ChargeReadout} and Fig.~\ref{fig:fig2}c), the projection results in the linear combination $\text{Twin}=0.92\cdot\text{SET}_{1} - 0.39\cdot\text{SET}_{2}$.

We note that the easiest form of correlation function is to implement a Twin SET with the modulated measurements. Upon demodulating each SET trace independently, the resulting scalars can be treated identically to a traditional measurement. The process of plotting the 2D histogram and projecting the data along the axis that maximizes the readout signal follows naturally from there, as shown in Fig.~\ref{fig:2D_hist_diff_mod}.

\begin{figure}[ht!]
\centering
\includegraphics[width=0.35\textwidth]{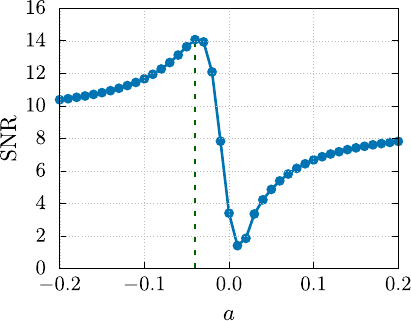}
\caption{SNR vs $a$ for the correlation function in Eq.~\ref{eq:corr} and the Rabi measurement from Fig.~\ref{fig:fig3}.}
\label{fig:prodOfffset_1D}
\end{figure}

The correlation function described in Eq.~\ref{eq:corr} has a free tuning parameter, $a$, which determines the weight between the product, $\Pi$, and the difference, $\Delta$, of the traces. The larger the value of $a$, the more weight is given to $\Delta$ and the suppression of uncorrelated noise enabled by $\Pi$ is not obtained. On the other hand, a value of $a$ close to zero will suppress $\Delta$, and the information about the anti-correlation will be lost. Consequently, finding the optimal value of $a$ is critical to obtain the best SNR from the correlation function. In Fig.~\ref{fig:prodOfffset_1D} we show the SNR fitting for the Rabi measurement described in Section~\ref{sec:SpinModulation} as a function of the value of $a$, and obtain an optimal value of $a = -0.04$. 

The process of finding the optimal $a$ is analogous to finding the optimal projection axis in the IQ plane, commonplace of any reflectometry technique. It is independent of integration time or modulation frequency (provided that the modulation frequency is above the $1/f$ corner), but dependent on the raw values of \SET1$(t)$ and \SET2$(t)$. As such, it needs to be retuned for every pair of dots, since the relative signals from \SET1 and \SET2 will vary. For each of the experiments described in this work, we present the SNR after finding the optimal value of $a$.

\section{Heralded Demodulation and Variance Transform}

A key advantage of measuring time traces instead of single integrated points is that a plethora of time-correlation functions can be studied. Moreover, the knowledge that the SETs will present predictable correlations in response to a charge transition opens a path for the study of unpredictable charge events. For instance, the study of charge noise in CMOS devices will benefit from the ability to identify, with high SNR, the time distribution of two-level fluctuators (TLF). Our proposed technique is ill-equipped for this task because applying a modulation tone filters out charge noise and suppresses TLF events. However, measuring time traces from a twin SET can properly serve this purpose by allowing one of the SETs to act as a herald that flags when a charge event has occurred. If a signal with the same time signature is also observed in the other SET, there is a high degree of certainty that it indeed corresponds to a charge event. Any local noise will be uncorrelated and the SNR for the correlated (or anti-correlated) signals will be enhanced.

A mathematical tool that is very well suited for this purpose is the variance transform \cite{Jiang2020_predictor}, where a wavelet decomposition of the signals is used to extract information about the time- and frequency-domain features of a pair of signals. The variance transform identifies the dominant features of the first signal and uses them to weigh the time and frequency components of the second signal. For a detailed description of how the variance transform can be used to identify correlation between pairs of experimental variables, see Refs.~\cite{Seedhouse2023_unpublished,Dumoulin2023_unpublished}.

\section{Accumulation of Decay Probability}\label{sec:decayAccum}

As the electrons are periodically moved between the dots, there is a non-zero probability of relaxation, even if the modulation period is much lower than the relaxation time, $T_{1}$. Indeed, it is the wait time $\tau_{0}$ at each charge configuration which allows for relaxation of the spin states towards the corresponding ground state. The probability that an arbitrary state $\ket{\phi}$ decays into a lower energy state $\ket{\psi}$ after a time $t$ is given by:
\begin{eqnarray}
P_{\textrm{decay}} = 1-e^{-t/T_{1}}
\end{eqnarray}
where $T_{1}$ is the relaxation time from $\ket{\phi}$ into $\ket{\psi}$.  After a wait time $\tau_{0}$ which corresponds to half the modulation period, the probability of preserving the spin state, i.e., that the state did not decay during that time, is:
\begin{eqnarray}
\bar{P}_{\textrm{decay}} = e^{-\tau_{0}/T_{1}}
\end{eqnarray}
By performing the periodic pulsing with $N$ periods, we are repeating the experiment $N$ independent times, provided that the state did not decay. As such, after $N$ modulation periods the cumulative probability that the spin has not decayed is:
\begin{eqnarray}
\bar{P}_{\textrm{decay}} = \left(e^{-\tau_{0}/T_{1}}\right)^{N}
\end{eqnarray}
from where it follows that the probability of decaying after the total integration time $\tau=2N\tau_{0}$ is:
\begin{eqnarray}
P_{\textrm{decay}} &=& 1-e^{-N\tau_{0}/T_{1}} \nonumber \\
P_{\textrm{decay}} &=& 1-e^{-\tau/ \left(2T_{1}\right)}
\end{eqnarray}

This result shows that the modulation effectively extends the relaxation lifetime by a factor of 2, since the state can only relax during the semi-period that it spends in the charge configuration where the relaxation takes place. On the other hand, pulsing to a different charge configuration opens a new channel through which the state can relax. This result also evidences that higher modulation frequencies (lower $\tau_{0}$) will require an increased number of periods to accumulate the same probability of decay. Consequently, operating at high $\fmod$ can lead to improved demodulation due to the larger number of periods whilst maintaining the state probability.

\section{Markov Model of Spin Dynamics}\label{sec:markovChain}

To construct the Markov model, we consider separately the process of crossing between the charge configurations and the relaxation that occurs while waiting at each of them. Here, we provide the phenomenological description with which we construct the model. Refer to the energy diagram in Fig.~\ref{fig:fig1}d) for a guide.

When considering the crossing process, we allow for spin flip errors to change the parity of the state. Going from $(3,1)$ to $(4,0)$, this means that, with probability $\Psf$, \uu\ will tunnel to \ud, \sdd\ to \singFZ, \singFZ\ to \uu, and \ud\ and \du\ to \sdd. Similarly, when crossing from $(4,0)$ to $(3,1)$ \uu\ is allowed to tunnel to \singFZ, \sdd\ to \du, \singFZ\ to \sdd, and \ud\ and \du\ to \uu. The error-less crossing is defined such that \du\ and \ud\ are indistinguishable from \singFZ, while the even parity states will be conserved.

For the relaxation process, we consider the five relaxation channels indicated in Table~\ref{table:T1s}. $\Pa$ ($\Pb$) is associated to the single-spin relaxation on the right (left) dot that does not involve dephasing.  Relaxation to \singFZ\ depends on the initial state and corresponds to the $T_{+}$, $T_{-}$ and $T_{0}$ state lifetimes \cite{Seedhouse2021}.

In the parity basis $\left(\singFZM,\ket{\uuM},\ket{\udM},\ket{\duM},\ket{\sddM}\right)^{T}$, the matrices corresponding to the Markov chain operators are:

\begin{widetext}		
	\begin{eqnarray}
	T_{1}^{4,0}=
	\begin{pmatrix}
	1 & \Pa\Pb\left(1-\Pg\right) & \Pa\left(1-\Pf\right) & \Pb\left(1-\Pf\right) & 1-\Pe \\
	0 & 1+3\Pa\Pb\Pg -\Pa\Pb-\Pa\Pg-\Pb\Pg & 0 & 0 & 0 \\
	0 & \Pa\Pg\left(1-\Pb\right) & 1+2\Pa\Pf-\Pa-\Pf & 0 & 0 \\
	0 & \Pb\Pg\left(1-\Pa\right) & 0 & 1+2\Pb\Pf-\Pb-\Pf & 0 \\
	0 & 0 & \Pf\left(1-\Pa\right) & \Pf\left(1-\Pb\right) & \Pe
	\end{pmatrix}
	\end{eqnarray}
\end{widetext}

	\begin{eqnarray}
	C_{3,1}^{4,0}=
	\begin{pmatrix}
	1-\Psf & 0 & 1-\Psf & 1-\Psf & \Psf \\
	\Psf & 1-\Psf & 0 & 0 & 0 \\
	0 & \Psf & 0 & 0 & 0 \\
	0 & 0 & 0 & 0 & 0 \\
	0 & 0 & \Psf & \Psf & 1-\Psf
	\end{pmatrix}
	\end{eqnarray}
	
	\begin{eqnarray}
	C_{4,0}^{3,1}=
	\begin{pmatrix}
	0 & \Psf & 1-\Psf & 1-\Psf & 0 \\
	0 & 1-\Psf & \Psf & \Psf & 0 \\
	1-\Psf & 0 & 0 & 0 & 0 \\
	0 & 0 & 0 & 0 & \Psf \\
	\Psf & 0 & 0 & 0 & 1-\Psf
	\end{pmatrix}
	\end{eqnarray}
	
	\begin{eqnarray}
	T_{1}^{3,1}=
	\begin{pmatrix}
	1 & 0 & 0 & 0 & 0 \\
	0 & 1+2\Pa\Pb -\Pa-\Pb & 0 & 0 & 0 \\
	0 & \Pa\left(1-\Pb\right) & \Pa & 0 & 0 \\
	0 & \Pb\left(1-\Pa\right) & 0 & \Pb & 0 \\
	0 & 0 & 1-\Pa & 1-\Pb & 1
	\end{pmatrix},
	\end{eqnarray}
		
where $\Psf$ is the probability that a spin flip occurs during the crossing and $\bar{P}_{i}=e^{-\tau_{0}/T_{1,i}}$ is the probability that a state does not relax after spending a time $\tau_{0}$ at the corresponding charge configuration, with $T_{1,i}$ the characteristic decay time. The time $\tau_{0}$ spent on each side of the charge transition is $\tau_{0}=1/2\fmod$. For the simulations presented in Fig.~\ref{fig:fig3}f), the characteristic $T_{1}$ times are indicated in Table~\ref{table:T1s}. We note that in this formalism, the state vectors do not correspond to wave-functions but to occupation probabilities of each of the basis states. As such, the matrices presented above are constructed to conserve the state probabilities (columns add up to 1), and no normalization is required. This allows us to adjust for the measurement visibility by setting the magnitude of the initial state vector. For the simulations presented in Fig.~\ref{fig:fig3} of the main text, the initial state vector is $0.95$\uu.

\begin{table}
\centering
\caption{Relaxation processes for the Markov chain model and their characteristic $T_{1}$ times.}
\begin{tabular}{cccc}\toprule
Parameter & Transition & $T_{1,i}$ (ms) & Configuration\\ \midrule
\multirow{2}{*}{$\Pb$} & \uu\ $\rightarrow$ \du & \multirow{2}{*}{10} & \multirow{2}{*}{(3,1) and (4,0)}\\ 
&  \ud\ $\rightarrow$ \sdd & & \\ \midrule
\multirow{2}{*}{$\Pa$} & \uu\ $\rightarrow$ \ud & \multirow{2}{*}{40} & \multirow{2}{*}{(3,1) and (4,0)}\\ 
&  \du\ $\rightarrow$ \sdd & & \\ \midrule
$\Pe$ & \sdd\ $\rightarrow$ \singFZ & 100 & (4,0)\\ \midrule
\multirow{2}{*}{$\Pf$} & \ud\ $\rightarrow$ \singFZ & \multirow{2}{*}{0.3} & \multirow{2}{*}{(4,0)}\\ 
&  \du\ $\rightarrow$ \singFZ & & \\ \midrule
$\Pg$ & \uu\ $\rightarrow$ \singFZ & 2 & (4,0) \\ \bottomrule
\end{tabular}
\label{table:T1s}
\end{table}

\begin{figure}
\centering
\includegraphics[width=0.49\textwidth]{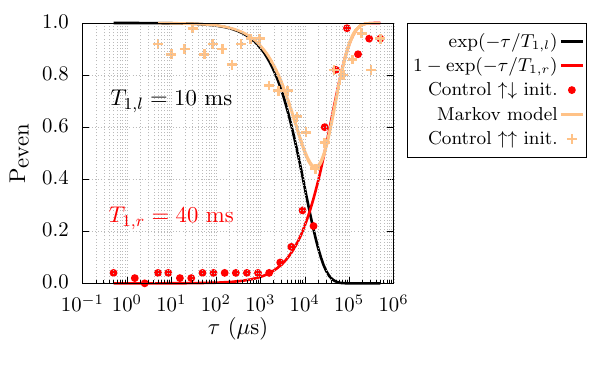}
\caption{Relaxation times for the control experiment. Data taken with a modulation amplitude of zero. The Markov model correctly reproduces the relaxation processes inside $(3,1)$.}
\label{fig:markovControl}
\end{figure}

In Fig.~\ref{fig:markovControl} we plot the results of the control experiment in which the modulation amplitude was set to zero for two different types of spin initialization: \uu\ and \ud. By fitting the decay rates, we obtain the values for $T_{1,l}$ and $T_{1,r}$ which we feed into the Markov model.

\section{Characterization of the Modulation Technique}\label{sec:modTechCharact}

\begin{figure}[ht!]
\centering
\includegraphics[width=0.42\textwidth]{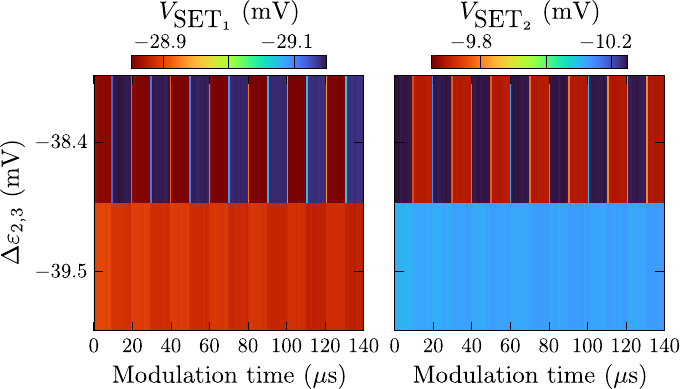}
\caption{Averaged time traces for SET$_{1}$ (left) and SET$_{2}$ (right) after applying the 2-point measurement on P2-P3. }
\label{fig:timeTrace_P2-P3_7periods}
\end{figure}

To help visualize the effect that the modulation has on the SETs, we show in Fig.~\ref{fig:timeTrace_P2-P3_7periods} the result of 100,000 averages of the SET traces after applying the 2-point measurement described in Sec.~\ref{sec:ChargeReadout}. The value of $\Delta\varepsilon_{2,3}=-38.4\ (-39.5)$ mV corresponds to the point close to (far from) the charge transition where the modulation causes (does not cause) the electron to tunnel.  As expected, the response of the SETs is much stronger when the electron actually tunnels across the dots, with the signal being anti-correlated between both SETs. The faint modulation present at $\Delta\varepsilon_{2,3}=-39.5$ mV is a consequence of the capacitive coupling that the modulation pulse itself has on the SETs.

\begin{figure}[ht!]
\centering
\includegraphics[width=0.4\textwidth]{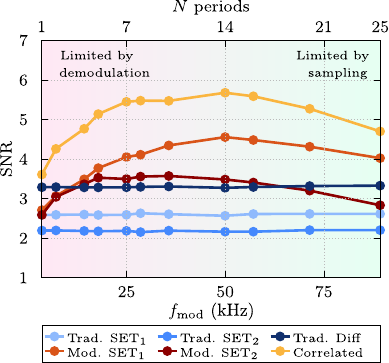}
\caption{SNR results for the P2-P3 system as a function of the modulation frequency and number of periods at constant integration time of 280 $\mu$s.}
\label{fig:snr_vs_fmod_P2P3}
\end{figure}

Figure~\ref{fig:snr_vs_fmod_P2P3} shows the SNR obtained for the modulation techniques at different modulations frequencies. The traditional methods are plotted for reference. At a constant integration time, $\tau=280\ \mu$s, the modulation frequency directly sets the number of periods of the SET traces. For lower frequencies, the number of periods is so low that the demodulation becomes less effective, thus reducing the SNR. As expected, the modulated results for both SETs tend towards the traditional results when the number of periods tends to one.

The decay of the SNR with increasing modulation frequency is a consequence of the limited sampling rate. Due to the memory limitations of the FPGA, the maximum sampling rate that was achieved to collect the full SET traces was 500 kSa/s. Nyquist theorem states that, in order to be reconstructed, the maximum bandwidth of a signal can be half of the sampling rate, 250 kHz in this case. Since we are modulating with a square wave composed of odd harmonics, the fundamental mode of a signal needs to be at most 83.3 kHz for the third harmonic to not undergo frequency folding.

\begin{figure}[ht!]
\centering
\includegraphics[width=0.49\textwidth]{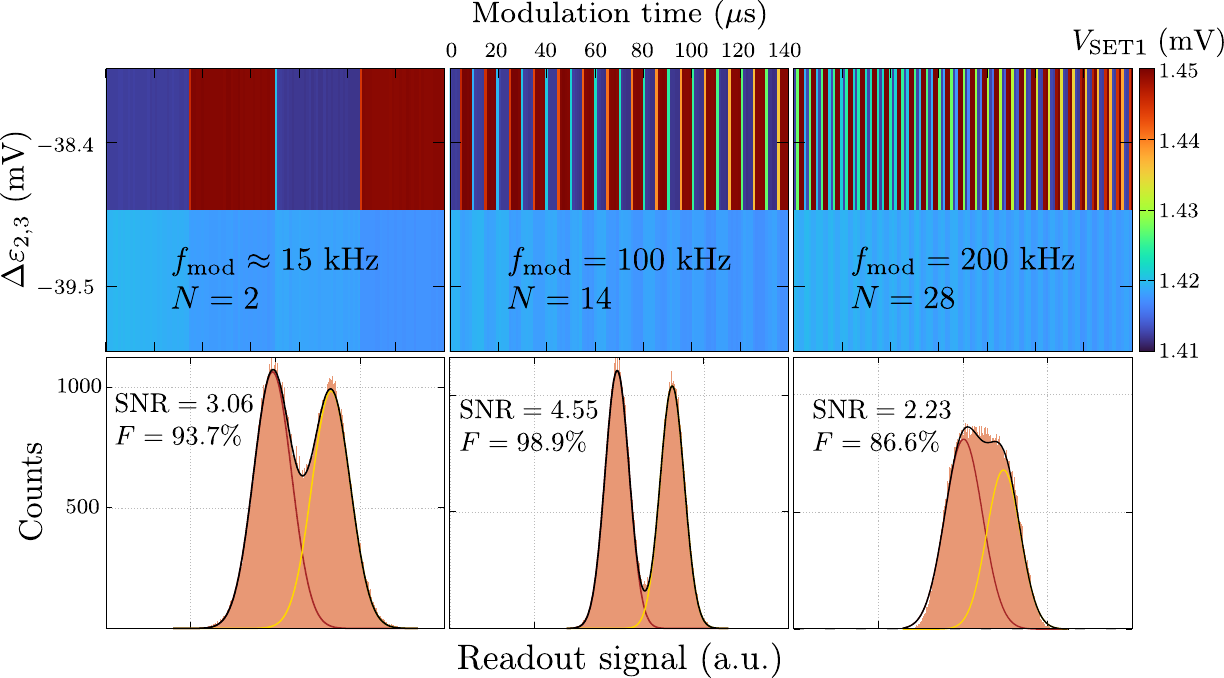}
\caption[Few-cycle demodulation and aliasing]{Few-cycle demodulation and aliasing. P2-P3 charge readout histograms for the modulated technique at different modulation frequencies. The total modulation time is kept constant at $\tau=140\ \mu$s. A few-cycle trace results in suboptimal demodulation and reduces the SNR (left panel). Similarly, when the modulation frequency becomes comparable to half of the sampling rate, aliasing occurs and the SNR drops (right panel). }
\label{fig:compareHistsPeriods}
\end{figure}

Figure~\ref{fig:compareHistsPeriods} shows the averaged SET traces and corresponding histograms after performing the single-electron measurement described in Section~\ref{sec:ChargeReadout} with a total integration time of 140 $\mu$s, and different modulations frequencies. For low $f_{\textrm{mod}}$ (2 periods), the histograms evidence a reduction in SNR due to the limited number of periods which hinders the demodulation. On the other hand, high $f_{\textrm{mod}}$ (28 periods) begin to show aliasing due to the limited sampling rate of 500 kSa/s, which also causes a reduction of the SNR. 

\begin{figure}[ht!]
\centering
\begin{subfigure}[h]{0.23\textwidth}
\includegraphics[width=0.9\textwidth]{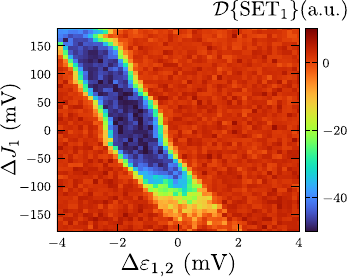}
\label{fig:detun_vs_J_P1-P2}
\end{subfigure}%
\begin{subfigure}[h]{0.23\textwidth}
\includegraphics[width=0.9\textwidth]{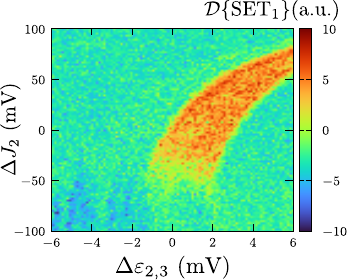}
\label{fig:detun_vs_J_P2-P3_SET1}
\end{subfigure}\\

\begin{subfigure}[h]{0.23\textwidth}
\includegraphics[width=0.9\textwidth]{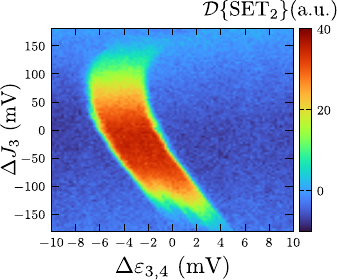}
\label{fig:detun_vs_J_P3-P4}
\end{subfigure}%
\begin{minipage}{0.23\textwidth}
		\hfill
\end{minipage}

\caption{Characterization of the modulation pulse with respect to the tunnel rate. Results of the demodulation as the detuning $\varepsilon_{i,j}$ is swept across the charge transition between dots $i$ and $j$. The tunnel coupling between the dots is controlled by $J_{i}$.}
\label{fig:ModCharacterization_J_modAmp}
\end{figure}

As seen in Fig.~\ref{fig:ModCharacterization_J_modAmp}, the result of the demodulation drops when the tunnel coupling (controlled by $J$) is reduced.  This is a direct consequence of the reduction in tunnel rate, which prevents the electron from tunnelling with the modulation. 

\section{Calculation of SNR and Readout Fidelity}

The histogram data is fitted to a bimodal gaussian function:
\begin{eqnarray}
G(x) &=& \dfrac{A_{1}}{\sqrt{2\pi}\sigma_{1}}\exp{-\dfrac{1}{2}\dfrac{(x-\mu_{1})^{2}}{\sigma_{1}^{2}}} \nonumber \\
&& \quad + \dfrac{A_{2}}{\sqrt{2\pi}\sigma_{2}}\exp{-\dfrac{1}{2}\dfrac{(x-\mu_{2})^{2}}{\sigma_{2}^{2}}}
\end{eqnarray}
where $\mu$, $\sigma$ and $A$ are, respectively, the mean, standard deviation and amplitude of each of the histogram peaks, and are left as free fitting parameters.

Once the distribution has been fitted, the SNR is calculated as:
\begin{eqnarray}
\text{SNR} = \dfrac{\left|\mu_{1}-\mu_{2}\right|}{\sqrt{\dfrac{1}{2}\left(\sigma_{1}^{2}+\sigma_{2}^{2}\right)}},
\end{eqnarray}
and the readout fidelity is then calculated as:
\begin{eqnarray*}
F = \dfrac{F_{1}+F_{2}}{2}
\end{eqnarray*}
where $F_{1,2}$ is the probability of correctly classifying a measurement in a given histogram peak (even/odd for spin readout; tunnel/no-tunnel for charge readout). The corresponding partial fidelities are:

\begin{eqnarray*}
F_{1} &=& \ddfrac{\int_{-\infty}^{th} e^{-\dfrac{1}{2}\dfrac{(x-\mu_{1})^{2}}{\sigma_{1}^{2}}} dx}{\int_{-\infty}^{\infty} e^{-\dfrac{1}{2}\dfrac{(x-\mu_{1})^{2}}{\sigma_{1}^{2}}} dx}, \\
\\
F_{2} &=& \ddfrac{\int_{th}^{\infty} e^{-\dfrac{1}{2}\dfrac{(x-\mu_{2})^{2}}{\sigma_{2}^{2}}} dx}{\int_{-\infty}^{\infty} e^{-\dfrac{1}{2}\dfrac{(x-\mu_{2})^{2}}{\sigma_{2}^{2}}} dx},
\end{eqnarray*}
where $th$ is the readout threshold, corresponding to the point where both gaussians intersect. We have assumed without loss of generality that $\mu_{1}<\mu_{2}$.

\bibliographystyle{apsrev}
\bibliography{references}

\end{document}